\documentclass[11pt]{article}
\usepackage{moriond,epsfig}

\def\Journal#1#2#3#4{{#1} {\bf #2}, #3 (#4)}

\def\NPB{{\em Nucl. Phys.} B}

\def\PRD{{\em Phys. Rev.} D}

\def\be{\begin{equation}}
\def\ee{\end{equation}}
\def\bea{\begin{eqnarray}}
\def\eea{\end{eqnarray}}

\begin{document}
\vspace*{4cm}
\title{Two-loop virtual QCD correction to the $B \rightarrow X_s l^+l^-$ decay}

\author{ ADRIAN GHINCULOV }

\address{Bausch and Lomb Hall, University of Rochester \\
         Rochester, NY 14627, USA }

\maketitle\abstracts{
I discuss an application of numerical techniques for massive two-loop Feynman diagram 
evaluation to a phenomenologically very important rare decay process, $B \rightarrow X_s l^+l^-$.
This process, currently being measured at B factories, has been calculated at two-loop level
in the lower dilepton invariant mass perturbative window by a previous collaboration. We extended
this result, by using numerical integration methods, to the whole observable spectrum of the
dilepton invariant mass consistent with the perturbative QCD regime.}


\section{Introduction}

B factories are opening the way to a class of precision electroweak measurements, including 
rare decays such as $B \rightarrow X_s l^+l^-$, which until very recently were beyond 
experimental reach. A measurement of the dilepton spectrum in this 
decay process has been proposed as a way to test for new physics and for discriminating 
between new physics scenarios. Even though the energy scale of this decay process is much lower
than the mass scale of the new physics being tested for, 
and therefore these additional heavy degrees of freedom
only appear as virtual effects, the $B \rightarrow X_s l^+l^-$ decay
is relatively sensitive to them because its leading order is one-loop.

This type of analysis relies both on an accurate measurement of the dilepton spectrum,
which is a matter of high integrated luminosity, and on an accurate theoretical treatment
of this process. Currently, the Monte Carlo simulators in use rely on leading order calculations.
One ingredient of an improved theoretical analysis is a calculation of two-loop QCD radiative
corrections; this will become increasingly necessary in the future, as the integrated 
luminosity increases.

However, to calculate the two-loop QCD radiative corrections to this process turns out to be
a difficult task. For a concise review of the theoretical effective Lagrangian framework involved,
see for instance ref. 1. The Bremsstrahlung process was recently calculated in refs. 1 and 2.
In their pioneering work of ref. 3, Asatryan {\em et al.} succeeded to calculate 
a mass and momentum
double expansion of the virtual two-loop corrections for this decay process. Because their
calculation is based on expansion techniques, this is a calculation of the lower dilepton 
spectrum perturbative window. Once the $c \bar c$ threshold is reached, the momentum expansion
is not a valid approximation anymore.

We have recently extended this calculation to the upper perturbative window -- see ref. 4. 
We stress that
experimentally this is an important kinematic zone where a comparable number of events 
are collected as in the lower perturbative window; however, one encounters 
nonperturbative corrections in this region. To do this, we resorted to numerical methods
which are valid both below and above the threshold. 
Regarding the lower perturbative window, our calculation serves as an independent confirmation
of ref. 3, which is particularly welcome in view of the complexity of this calculation and of the
fact that ref. 3 works in the mass and momentum expansion approximation.
We found good agreement with 
Asatryan {\em et al.}
at low dilepton invariant mass, while extending the result above the threshold. A detailed 
description of the calculation, the inclusion of the available Bremsstrahlung along with 
the virtual corrections, and a phenomenological analysis of the results, will 
be reported in ref. 4. Here we describe the calculation of the virtual QCD corrections.

\section{The calculation}

The relevant two-loop Feynman diagrams are shown in figure 1 . They can be organized in 
five gauge invariant subsets. This is useful because gauge cancellations occur within each 
subgroup, and gauge invariance for each subset is a useful check on the calculation.

As a first step toward the calculation of these diagrams, they are processed with a 
computer algebra program. We used two independent versions written in FORM and in Schoonship,
which provides a powerful check on the algebra. The aim of the algebraic manipulations
is to reduce the diagrams to a standard form which can be further integrated numerically.
The steps involved are the Dirac algebra, introducing Feynman parameters to write the diagrams
as multiple integrals over sunset-type tensor functions, Lorentz decomposition of the tensor
structures and isolation of the scalar integrals, and use of differential recursion relations
to reduce the scalar functions to a set of ten master scalar functions -- see ref. 5. As 
detailed in ref. 5, the crucial point here is that any two-loop diagram in renormalizable theories
can be decomposed by this algorithm into an expression involving only this limited set of ten
scalar integrals. This makes this algorithm applicable, given enough computing power, to any
two-loop process (up to possible infrared issues which, when manageable, must be handled 
analytically by hand).

After this algebra step, the Feynman diagrams are expressed as multiple integrals 
over a set of ten functions $h_i$ (or possibly their derivatives), where $h_i$ are the UV finite
parts of a set of sunset-type scalar functions. 
If we denote the three propagator masses of the sunset-type functions by $m_1$, $m_2$, and $m_3$, 
and the external momentum by $k$, the ten scalar functions $h_i$ are defined by the 
following integral representations:

\begin{eqnarray}
    h_1(m_1,m_2,m_3;k^2) & = &  \int_0^1 dx \,
                                \tilde{g} (x)
  \nonumber \\
    h_2(m_1,m_2,m_3;k^2) & = &  \int_0^1 dx \,
                              [ \tilde{g}   (x)
                              + \tilde{f_1} (x) ]
  \nonumber \\
    h_3(m_1,m_2,m_3;k^2) & = &  \int_0^1 dx \, 
                              [ \tilde{g}   (x)
                              + \tilde{f_1} (x) ] \, (1-x)
  \nonumber \\
    h_4(m_1,m_2,m_3;k^2) & = &  \int_0^1 dx \,
                              [ \tilde{g}   (x)
                              + \tilde{f_1} (x)
                              + \tilde{f_2} (x) ]
  \nonumber \\
    h_5(m_1,m_2,m_3;k^2) & = &  \int_0^1 dx \,
                              [ \tilde{g}   (x)
                              + \tilde{f_1} (x)
                              + \tilde{f_2} (x) ] \, (1-x)
  \nonumber \\
    h_6(m_1,m_2,m_3;k^2) & = &  \int_0^1 dx \,
                              [ \tilde{g}   (x)
                              + \tilde{f_1} (x)
                              + \tilde{f_2} (x) ] \, (1-x)^2
  \nonumber \\
    h_7(m_1,m_2,m_3;k^2) & = &  \int_0^1 dx \,
                              [ \tilde{g}   (x)
                              + \tilde{f_1} (x)
                              + \tilde{f_2} (x)
                              + \tilde{f_3} (x) ]
  \nonumber \\
    h_8(m_1,m_2,m_3;k^2) & = &  \int_0^1 dx \,
                              [ \tilde{g}   (x)
                              + \tilde{f_1} (x)
                              + \tilde{f_2} (x)
                              + \tilde{f_3} (x) ] \, (1-x)
  \nonumber \\
    h_9(m_1,m_2,m_3;k^2) & = &  \int_0^1 dx \,
                              [ \tilde{g}   (x)
                              + \tilde{f_1} (x)
                              + \tilde{f_2} (x)
                              + \tilde{f_3} (x) ] \, (1-x)^2
  \nonumber \\
    h_{10}(m_1,m_2,m_3;k^2) & = &  \int_0^1 dx \,
                              [ \tilde{g}   (x)
                              + \tilde{f_1} (x)
                              + \tilde{f_2} (x)
                              + \tilde{f_3} (x) ] \, (1-x)^3
   \; \; .
\label{eq:intrepr}
\end{eqnarray}

where we used the following notations:

\begin{eqnarray}
  \tilde{g} (m_1,m_2,m_3;k^2;x) & = &
     Sp(\frac{1}{1-y_1}) 
   + Sp(\frac{1}{1-y_2}) 
   + y_1 \log{\frac{y_1}{y_1-1}} 
   + y_2 \log{\frac{y_2}{y_2-1}} 
  \nonumber \\
  \tilde{f_1}(m_1,m_2,m_3;k^2;x) & = &
   \frac{1}{2}
   \left[
   - \frac{1-\mu^2}{\kappa^2}
   + y_1^2 \log{\frac{y_1}{y_1-1}} 
   + y_2^2 \log{\frac{y_2}{y_2-1}} 
   \right]
  \nonumber \\
  \tilde{f_2}(m_1,m_2,m_3;k^2;x) & = &
   \frac{1}{3}
   \left[
   - \frac{2}{\kappa^2} 
   - \frac{1-\mu^2}{2 \kappa^2}
   - \left( \frac{1-\mu^2}{\kappa^2} \right)^2
   \right.
  \nonumber \\
 & &
   \; \; \; \; \; \; \; \; \; \; \; \;
   \; \; \; \; \; \; \; \; \; \; \; \;
   \; \; \; \; \; \; \; \; \; \; \; \;
   \left.
   + y_1^3 \log{\frac{y_1}{y_1-1}} 
   + y_2^3 \log{\frac{y_2}{y_2-1}} 
   \right]
  \nonumber \\
  \tilde{f_3}(m_1,m_2,m_3;k^2;x) & = &
   \frac{1}{4}
   \left[
   - \frac{4}{\kappa^2} 
   - \left( \frac{1}{3} + \frac{3}{\kappa^2}  \right) 
     \left( \frac{1-\mu^2}{\kappa^2} \right)
   - \frac{1}{2} \left( \frac{1-\mu^2}{\kappa^2} \right)^2
   - \left( \frac{1-\mu^2}{\kappa^2} \right)^3
   \right.
  \nonumber \\
 & &
   \left.
   \; \; \; \; \; \; \; \; \; \; \; \;
   \; \; \; \; \; \; \; \; \; \; \; \;
   \; \; \; \; \; \; \; \; \; \; \; \;
   + y_1^4 \log{\frac{y_1}{y_1-1}} 
   + y_2^4 \log{\frac{y_2}{y_2-1}} 
   \right]
   \; \; ,
\end{eqnarray}
%
\begin{eqnarray}
y_{1,2} & = & \frac{1 + \kappa^{2} - \mu^{2}
                    \pm \sqrt{\Delta}}{2 \kappa^{2}}  \nonumber \\
\Delta  & = & (1 + \kappa^{2} - \mu^{2})^{2} 
          + 4 \kappa^{2} \mu^{2} - 4 i \kappa^{2} \eta 
      \; \; ,
\end{eqnarray}
%
\begin{eqnarray}
   \mu^{2}  & = &  \frac{a x + b (1-x)}{x (1-x)}   \nonumber \\
         a  & = &  \frac{m_{2}^{2}}{m_{1}^{2}} \, , \; \; \; \;
         b \; = \; \frac{m_{3}^{2}}{m_{1}^{2}} \, , \; \; \; \;
\kappa^{2} \; = \; \frac{    k^{2}}{m_{1}^{2}} 
      \; \; .
\end{eqnarray}

\begin{figure}
\begin{center}
a) \psfig{figure=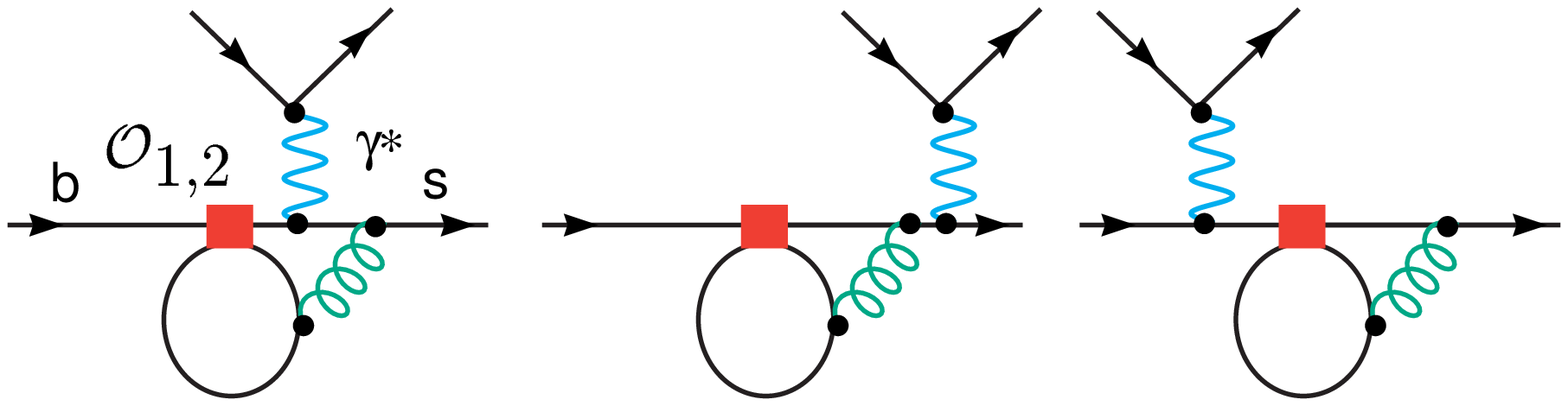,height=1.5in}

b) \psfig{figure=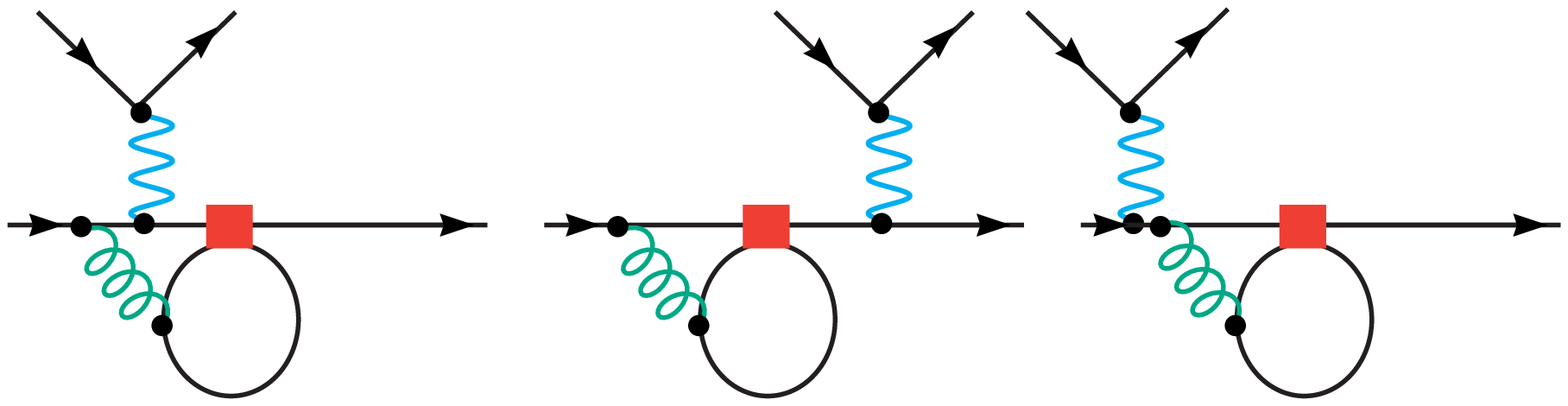,height=1.5in}

c) \psfig{figure=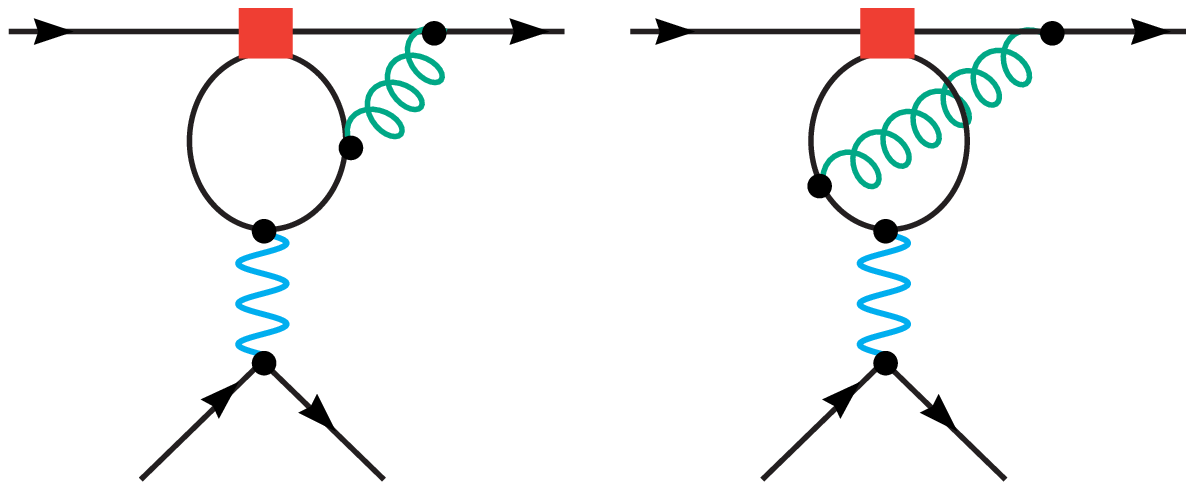,height=1.5in}

d) \psfig{figure=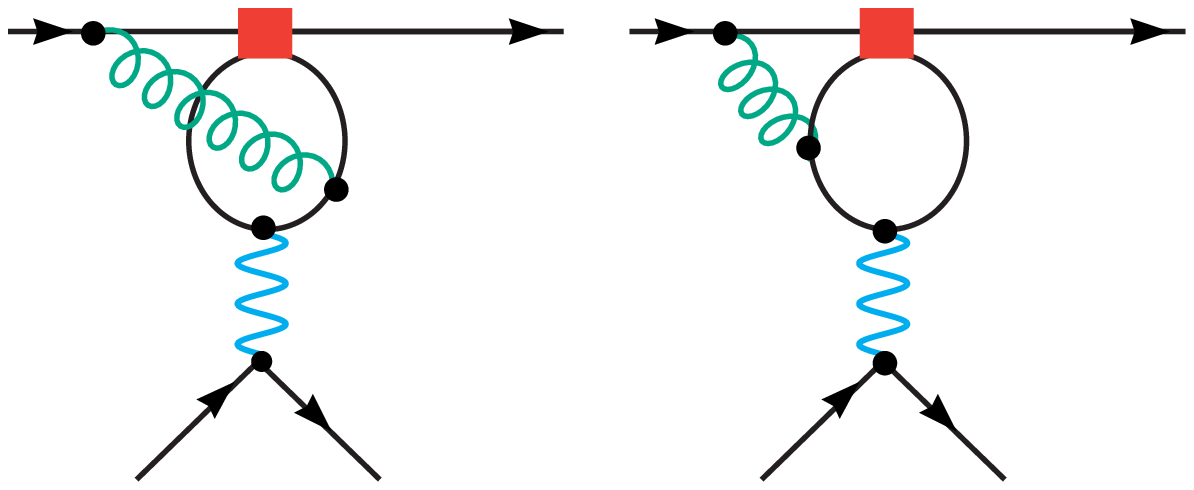,height=1.5in}

e) \psfig{figure=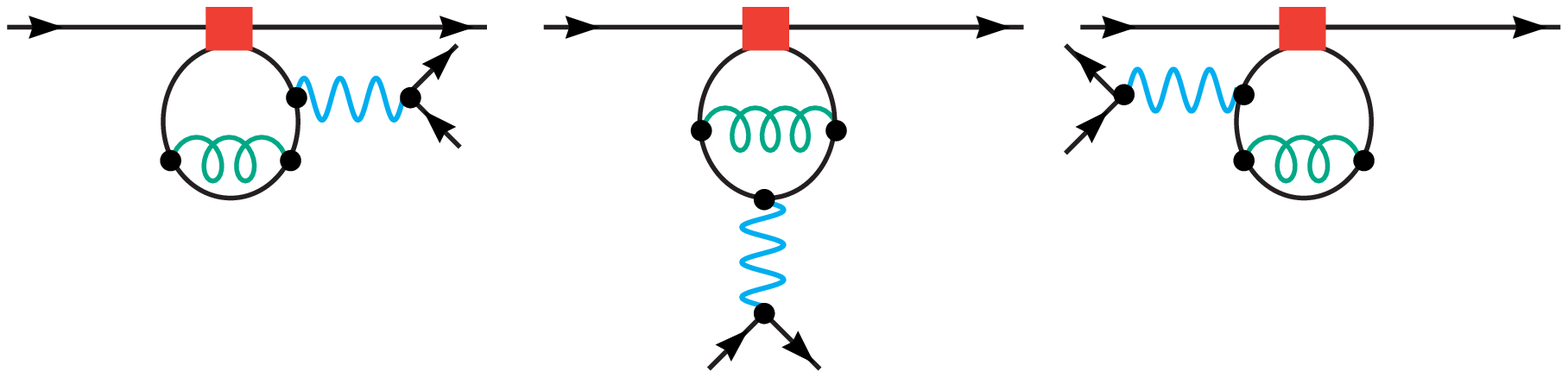,height=1.5in}
\end{center}
\caption{Two-loop Feynman diagrams relevant for the virtual QCD corrections
to the rare decay $B \rightarrow X_s l^+l^-$. They can be organized in five gauge 
invariant subsets.
\label{fig:radish}}
\end{figure}

\begin{figure}
\vskip 5.5cm
\psfig{figure=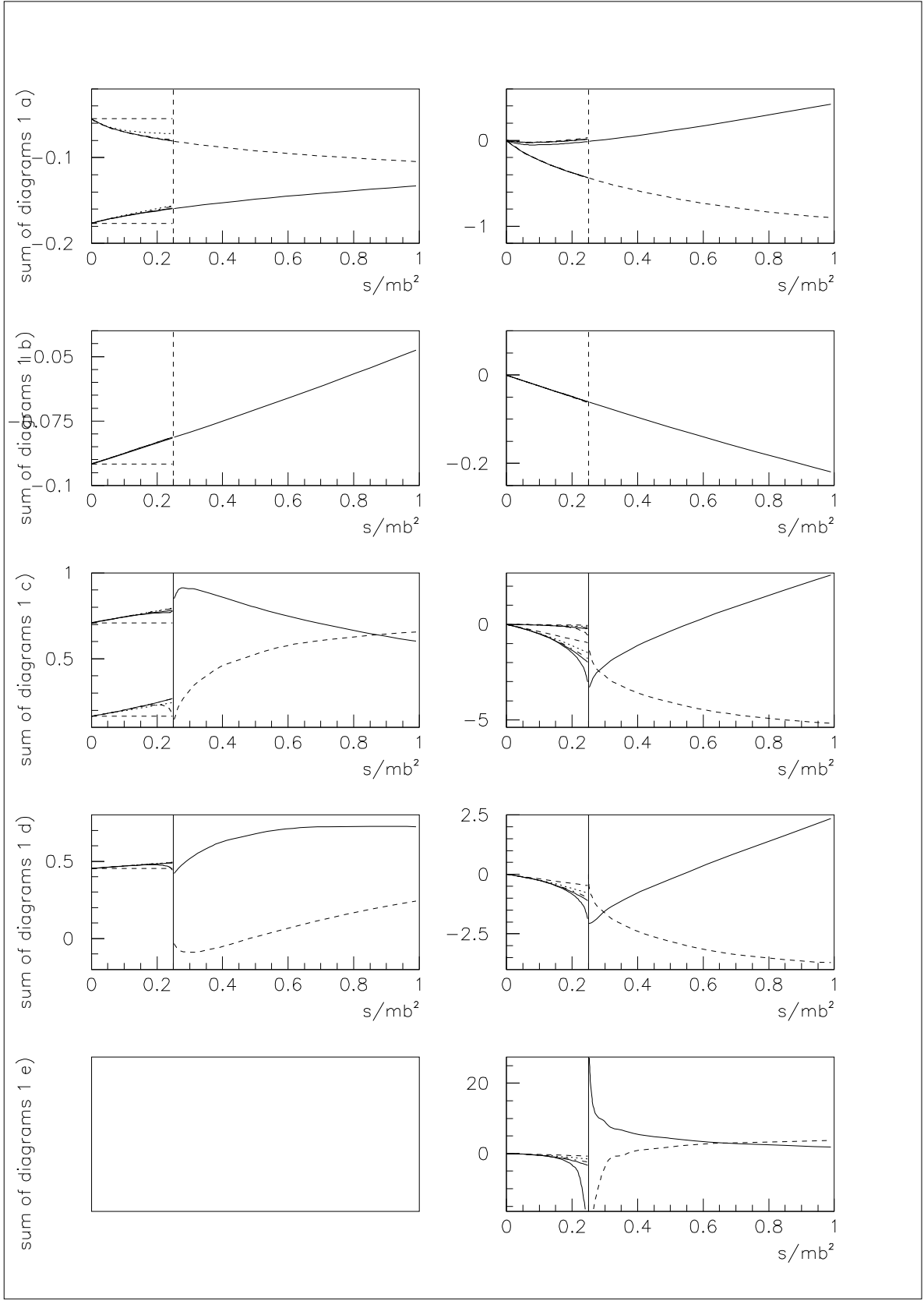,height=6.5in}
\caption{Plots of the UV finite part of the
Feynman diagram subsets shown in figure 1. The two columns correspond to the electric and the
magnetic form factors. We plot the real and the imaginary parts of our exact numerical
integration result, along with successive approximations in the momentum expansion series
of the result of Asatryan {\em et al.} which is shown only below the $c\bar c$ threshold.
\label{fig:radish}}
\end{figure}

\begin{figure}
\vskip 5.5cm
\psfig{figure=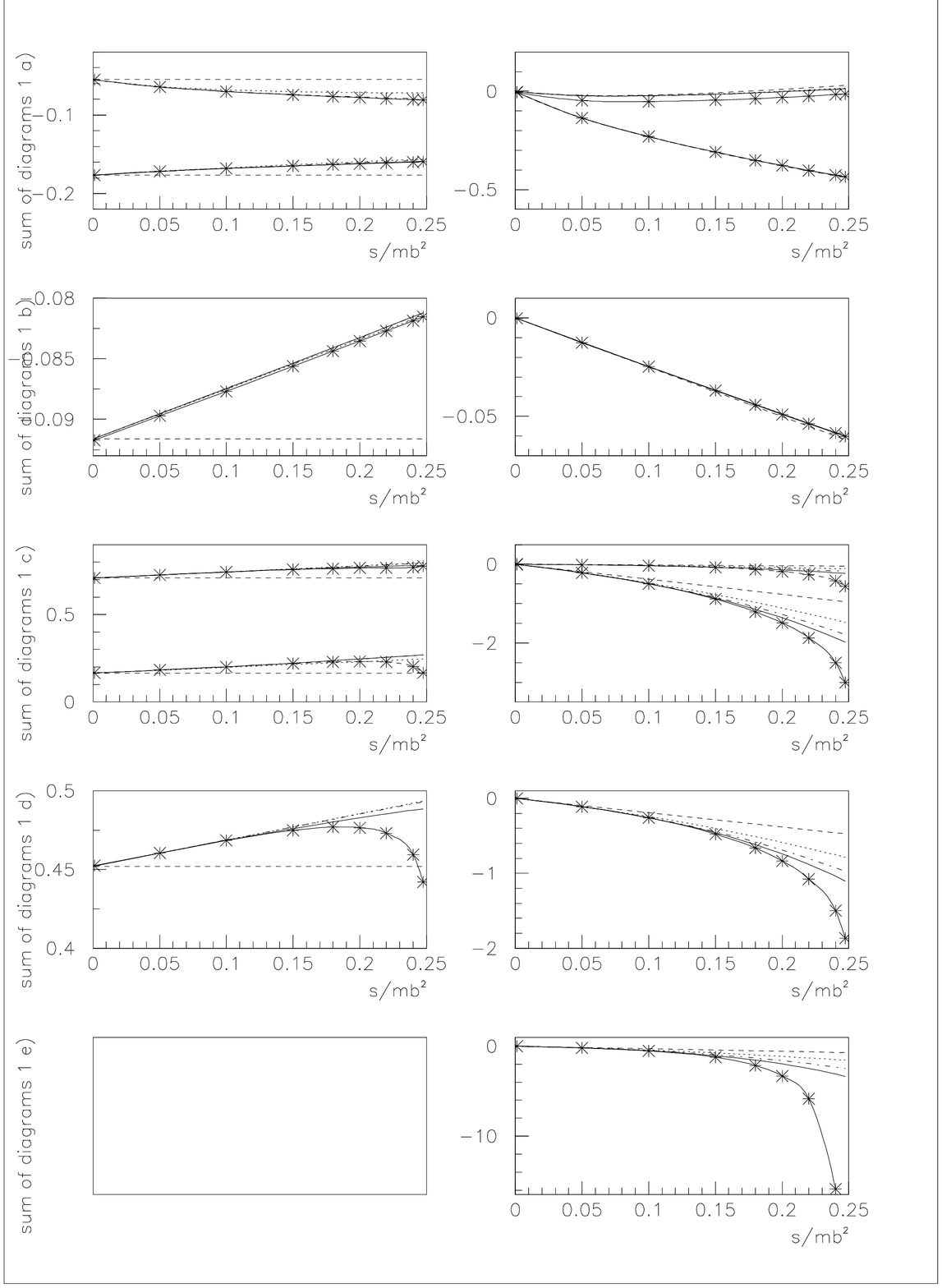,height=6.5in}
\caption{
Convergence pattern of the momentum expansion solution of Asatryan {\em et al.}
toward our numerical integration solution. The stars denote actual integration points, and the 
solid line connecting them is an interpolation.
\label{fig:radish}}
\end{figure}

The analytical expressions that result after this first step need to be further 
integrated numerically. We use an adaptive deterministic integration algorithm,
which is well suited for accurate numerical integrations when the dimensionality of the
integral is not too large. In the problem at hand, we deal with three-fold numerical integrations
at most. We note that four-fold numerical integrations were also shown to be feasible within
the context of our two-loop numerical techniques in ref. 6. Based on the computing time that
the four-fold integrations take, it appears likely that in the case of two-loop topologies 
more complicated than three-point functions, where numerical integration of higher dimensionality 
than four may occur, it will become more practical to use other numerical integration 
methods that become superior in higher dimensions, such as number-theoretical integration methods.

In general, the integrals over Feynman parameters must be
performed along a complex integration path that is consistent with the causality
condition. This path is computed automatically, by using spline functions. 
We refer to refs. 5---7 for details about the numerical integration.

In the case of the $b \rightarrow s l^+l^-$ decay, we deal with three kinematic variables:
the charm mass, the dilepton invariant mass, and the subtraction scale, all normalized by the 
mass of the bottom quark. In order for the result to be usable for phenomenological studies,
in particular to be implemented in a Monte Carlo simulation, we need to cover this 
three dimensional kinematic space. The actual calculation of the two-loop matrix element
by numerical integration is far too slow to be implemented directly into a Monte Carlo
simulation of the experimental setup. Therefore, a practical solution is to set up a suitable
grid of points that cover the whole three dimensional kinematic range, and to interpolate 
between the points. This results into a fast program that calculates the two-loop matrix element,
and which is fast enough to be included into a Monte Carlo simulation.

We selected a grid of $38 \times 3 \times 3$ integration points for both the electric 
and the magnetic components of the two-loop virtual correction. This leads to 684 integration 
points for the form factors involved in this calculation (see ref. 4). Each 
integration point was calculated with a relative precision of $10^{-3}$. We used the CERN
Linux cluster to perform this calculation, and the CPU usage was approximately 3 days 
on 33 processors (mostly 850 MHz) running in parallel. 
It is obvious that with the present-day
processors, calculating in advance a comprehensive grid of integration points, storing them, and 
then interpolating between them is the practical way to calculate such radiative corrections 
efficiently.

\section{Results}

In figure 2 we show the results for each of the five gauge invariant diagram subsets. We show
both our results, valid over the whole range of the dilepton invariant mass, compared to the 
results of Asatryan {\em et al.}, 
valid in principle only below the $c \bar c$ threshold. In figure 3
we compare the way how the momentum expansion converges toward our exact numerical 
solution \footnote{We thank M. Walker for providing us with partial results of the
calculation described in ref. 3, which made possible this detailed comparison 
with our results.}.

There is good agreement between our results for each diagram set and the double expansion, 
which provides a strong test of our solution.
One notices that, as a general rule, the less singular the threshold behaviour of the diagram,
the better the momentum expansion converges toward our exact numerical result. In the case of the
gauge invariant subset e) we notice the poorest convergence of the momentum expansion. This is
correlated with the singular behaviour of this function at threshold. We note in passing that
this singular behaviour is due chiefly to the charm self-energy type diagrams; this singularity
is subsequently canceled when the counterterm contribution to the charm self-energy is added.
This makes the agreement between our final physical result for the radiative corrections and 
the momentum expansion result to be better than what can be inferred from figure 3 alone.

\section{Conclusions}

To conclude, we calculated the virtual two-loop QCD corrections to the rare decay
process $B \rightarrow X_s l^+l^-$. Our calculation is valid over the whole dilepton 
spectrum that is observable, over both the lower and the upper perturbative windows 
separated by the $c\bar c$ resonance zone. In the low dilepton invariant mass limit, 
we found very good agreement with an existing result by Asatryan {\em et al.}, 
which resorts to a double expansion in the dilepton invariant mass 
and in the charm mass. This provides a powerful check of our result. The complete NNLO
QCD effect, which includes the Bremsstrahlung along with the virtual corrections,
and the phenomenological implications of these QCD corrections, will be discussed in a 
forthcoming article [4].

\section*{Acknowledgments}
This work was supported by the US Department of Energy (DOE).
Special thanks are due to Tobias Hurth (CERN), Gino Isidori (Frascati), and
York-Peng Yao (University of Michigan) for the enjoyable 
collaboration on which this talk is based.
I am grateful to the CERN Theory Division for hospitality during several visits when 
parts of this work were completed.

\end{document}